\documentclass[sigconf]{acmart}

\setcopyright{none}
\renewcommand\footnotetextcopyrightpermission[1]{}
\newtheorem{myDef}{Definition}
\usepackage{subfigure}
\begin{document}

\title{Harnessing Large Language Models for Text-Rich \\Sequential Recommendation}

\author{Zhi Zheng}
\affiliation{%
  \institution{School of Data Science, University of Science and Technology of China}
  \country{China}}
\affiliation{%
  \institution{Career Science Lab, BOSS Zhipin}
  \country{China}}
\email{zhengzhi97@mail.ustc.edu.cn}

\author{Wenshuo Chao}
\affiliation{%
  \institution{Hong Kong University of Science and Technology (Guangzhou)}
  \country{China}}
\affiliation{%
  \institution{Career Science Lab, BOSS Zhipin}
  \country{China}}
\email{wchao829@connect.hkust-gz.edu.cn}

\author{Zhaopeng Qiu}
\affiliation{%
  \institution{Career Science Lab, BOSS Zhipin}
  \country{China}}
\email{zhpengqiu@gmail.com}

\author{Hengshu Zhu}
\authornote{Corresponding authors.}
\affiliation{%
  \institution{Career Science Lab, BOSS Zhipin}
  \country{China}}
\affiliation{%
  \institution{Thrust of Artificial Intelligence, Hong Kong University of Science and Technology (Guangzhou)}
  \country{China}}
\email{zhuhengshu@gmail.com}

\author{Hui Xiong}
\authornotemark[1]
\affiliation{%
  \institution{Thrust of Artificial Intelligence, Hong Kong University of Science and Technology (Guangzhou)}
  \country{China}}
\affiliation{%
  \institution{Department of Computer Science and Engineering, Hong Kong University of Science and Technology}
  \country{Hong Kong SAR}}
\email{xionghui@ust.hk}


\begin{abstract}
Recent advances in Large Language Models (LLMs) have been changing the paradigm of Recommender Systems (RS). However, when items in the recommendation scenarios contain rich textual information, such as product descriptions in online shopping or news headlines on social media, LLMs require longer texts to comprehensively depict the historical user behavior sequence. This poses significant challenges to LLM-based recommenders, such as over-length limitations, extensive time and space overheads, and suboptimal model performance. To this end, in this paper, we design a novel framework for harnessing Large Language Models for Text-Rich Sequential Recommendation (LLM-TRSR). Specifically, we first propose to segment the user historical behaviors and subsequently employ an LLM-based summarizer for summarizing these user behavior blocks. Particularly, drawing inspiration from the successful application of Convolutional Neural Networks (CNN) and Recurrent Neural Networks (RNN) models in user modeling, we introduce two unique summarization techniques in this paper, respectively hierarchical summarization and recurrent summarization. Then, we construct a prompt text encompassing the user preference summary, recent user interactions, and candidate item information into an LLM-based recommender, which is subsequently fine-tuned using Supervised Fine-Tuning (SFT) techniques to yield our final recommendation model. We also use Low-Rank Adaptation (LoRA) for Parameter-Efficient Fine-Tuning (PEFT). We conduct experiments on two public datasets, and the results clearly demonstrate the effectiveness of our approach. 
\end{abstract}

\maketitle
\section{INTRODUCTION}
\label{sec:intro}
Recently, Large Language Models (LLM), exemplified by ChatGPT\footnote{https://chat.openai.com/}, have demonstrated remarkable capabilities in the field of Natural Language Processing (NLP), capturing the attention of numerous researchers. Owing to the strong reasoning and zero/few-shot learning capabilities exhibited by LLMs, many researchers are also exploring their application in other domains, such as Recommender Systems (RS)~\cite{wu2023survey}. According to Wu et al.~\cite{wu2023survey}, a typical paradigm for employing LLM as RS involves feeding user profiles, behavioral data, and task instruction into the model, with the expectation that the LLM will offer a reasonable recommendation result in return. For example, Bao et al.~\cite{bao2023tallrec} propose TALLRec, which converts the history sequence and new item to "Rec Instruction" and "Rec Input" as the input for the LLM model.

However, in recommendation scenarios where items have rich textual information, e.g., product titles in e-commerce, news headlines on media platforms, extended text becomes essential to comprehensively depict a user historical behavior sequence, which introduces the following challenges to LLMs. First, existing LLMs typically impose limitations on the length of the input, e.g., 1,024 tokens for GPT-2~\cite{radford2019language}, which may be insufficient to encompass extensive textual information. Second, due to the $O(n^2)$ computational complexity of the Transformer~\cite{vaswani2017attention} architecture, prolonged texts lead to significant computational resource overheads for downstream recommendation tasks, which poses challenges to applications of recommender systems that demand high real-time responsiveness. Third, lengthier texts can make it more challenging for the model to effectively capture shifts in user preferences, potentially hindering optimal performance~\cite{liu2023lost}.

\begin{figure}[tb]
    \centering
    \includegraphics[width=0.9\linewidth]{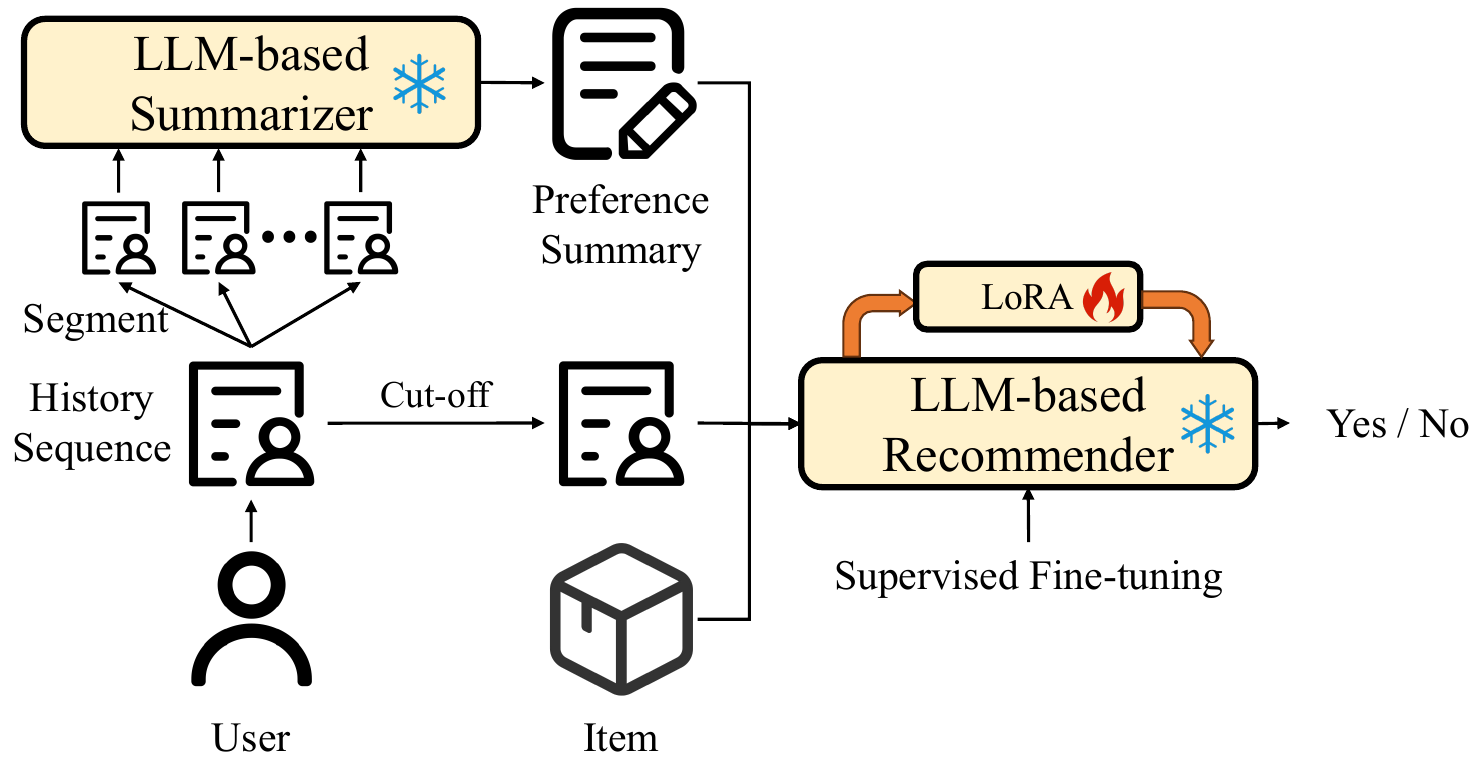}
    \vspace{-2mm}
    \caption{A schematic diagram of our method. The blue frost symbol indicates fixed parameters, while the red flame symbol signifies parameters that are updated during training.}
    \label{fig:diagram}
    \vspace{-2mm}
  \end{figure}

To this end, in this paper, we design a novel framework for harnessing Large Language Models for Text-Rich Sequential Recommendation (LLM-TRSR). Figure~\ref{fig:diagram} shows a schematic diagram of our proposed method. Specifically, our method mainly consists of the following primary steps. Initially, we extract the user behavioral history sequence and transform it into an extended piece of text. Subsequently, this long text is segmented into several blocks, ensuring that each block can be fully ingested by large language models. We then propose an LLM-based summarizer that holistically considers these blocks to derive a summary of user preference. Note that the parameters of this summarizer are frozen. Finally, we build the input prompt text based on the user preference summary, recent user interactions, and candidate item information, and feed the prompt into an LLM-based recommender,  which is subsequently fine-tuned using Supervised Fine-Tuning (SFT) to output "Yes" or "No". Additionally, a Parameter-Efficient Fine-Tuning (PEFT) method based on Low-Rank Adaptation (LoRA)  to reduce memory overhead and expedite the training process.

In the aforementioned process, a pivotal question is how to employ an LLM-based summarizer to extract user preference from multiple blocks of user behavior. In this paper, inspired by two neural network architectures which are extensively applied in the deep learning domain, respectively Convolutional Neural Network (CNN) and Recurrent Neural Network (RNN), we propose two distinct summarization approaches, respectively hierarchical summarization and recurrent summarization. Specifically, for the hierarchical summarization paradigm, we first employ the LLM-based summarizer to extract a summary from each individual block. Subsequently, these individual summaries are concatenated progressively and input to the summarizer again, leading to a higher-level summarization. Through this hierarchical approach, we achieve the final preference summary of the user. For the recurrent summarization paradigm, we initiate the process by using the LLM-based summarizer to extract a summary from the first block. Following this, we iteratively feed the subsequent blocks along with the previously generated summary back into the LLM-based summarizer, prompting it to update the user preference summary based on the new behavioral input. This iterative process continues until the final block, culminating in a comprehensive preference summary of user behavior.

To demonstrate the effectiveness of our method, we conduct extensive experiments on two publicly available datasets from distinct domains, respectively the Amazon-M2 dataset~\cite{jin2023amazon} tailored for product recommendation in e-commerce, and the MIND dataset~\cite{wu2020mind} designed for news recommendations on media platforms. The experimental results clearly demonstrate the efficacy of our approach. The major contributions of this paper can be summarized as:
\begin{itemize}
    \item To the best of our knowledge, we are the first to propose harnessing large language models to address the text-rich sequential recommendation problem.
    \item We propose to utilize an LLM-based summarizer to encapsulate user behavioral history, and we introduce two distinct summarization paradigms, respectively hierarchical summarization and recurrent summarization.
    \item We validated the effectiveness of our approach on two open-source datasets from distinct domains. The code is available at https://github.com/zhengzhi-1997/LLM-TRSR.
\end{itemize}
\section{RELATED WORK}
In this section, we will summarize the related works in the following three categories, respectively sequential recommendation, large language models, and LLM for recommendation.
\subsection{Sequential Recommendation}
Sequential recommender systems, as a specialized form of recommendation system~\cite{wang2020setrank,zheng2021drug,zheng2023interaction,zheng2022cbr,zheng2022ddr}, has garnered substantial attention in recent years due to their significance in modeling user dynamic behaviors and interests~\cite{chang2014predicting,zhang2021talent,wang2021variable,han2023guesr,qin2023comprehensive,hu2023boss}. Existing sequential recommendation models primarily employ sequence modeling techniques such as RNN or Transformer to represent user behavior sequences. For example, GRU4Rec~\cite{hidasi2015session} proposes to leverage the Gated Recurrent Unit (GRU) model for session-based data modeling. NARM~\cite{li2017neural} further explores a hybrid encoder with an attention mechanism to capture the user purpose in the current session. BERT4Rec~\cite{sun2019bert4rec} leverages the BERT-based deep bidirectional self-attention architecture to get the representation of user behavior sequences. Furthermore, items in recommender systems might also encompass abundant side information, especially textual data, e.g., the title of news or products. Therefore, some studies have incorporated additional modules for text-rich sequential recommendation scenarios. For example, LSTUR~\cite{an2019neural} designs a news encoder based on CNN and attention to get the new embedding, and further leverages GRU for sequential modeling. TempRec~\cite{wu2022news} also designs an item encoder and utilizes Transformer for sequential modeling. However, the aforementioned studies have not extensively explored the utilization of LLMs in sequential recommendation.

\subsection{Large Language Models}
Large Language Models are advanced linguistic models consisting of neural networks ranging from tens of millions to trillions of parameters, trained substantially on vast volumes of untagged texts using methods like self-supervised or semi-supervised learning approaches~\cite{min2021recent,zhao2023survey}. The foundation for these LLMs is the Transformer~\cite{vaswani2017attention} structure, which stands as a cornerstone in the field of deep learning for Natural Language Processing (NLP). Typically, LLMs can be classified into two different types, respectively discriminative LLMs and generative LLMs. For discriminative LLMs, BERT~\cite{kenton2019bert} introduces a bidirectional transformer architecture and establishes the concept of the Masked Language Model (MLM) for model pre-training. XLNet~\cite{yang2019xlnet} incorporates sequence order permutations, facilitating comprehension of word contexts within their surrounding lexical environment. For the generative LLMs, GPT~\cite{radford2018improving} first proposes to pre-train the model by predicting the next word in a sentence. InstructGPT~\cite{ouyang2022training} further proposes Reinforcement Learning from Human Feedback (RLHF) for fine-tuning. Llama and Llama-2~\cite{touvron2023llama, touvron2023llama2} are two famous collections of LLMs ranging in scale from 7 billion to 70 billion parameters. In this paper, we select the Llama models as the summarizer and recommender. Recently, several studies have focused on how to extend the input length limitations of existing LLMs~\cite{xiong2023effective}. However, the challenges of increased computational overhead and performance degradation remain unresolved. Through the user preference summarization method proposed in this paper, we can handle theoretically infinite user behavior sequences and significantly reduce the training overhead of downstream recommendation models.

\subsection{LLM for Recommendation}
Due to the powerful reasoning capabilities and zero/few-shot learning abilities, LLMs have recently gained significant attention in the domain of recommender systems~\cite{fang2023recruitpro,wu2023exploring,zheng2023generative}. According to the survey paper~\cite{wu2023survey}, existing studies on LLM for RS can be divided into two categories, respectively discriminative LLMs for RS and generative LLMs for RS, and the modeling paradigm can be divided into three categories, respectively LLM Embeddings + RS, LLM Tokens + RS, and LLM as RS. Indeed, the discriminative LLM for recommendation mainly refers to the BERT-based models, while the generative LLM for recommendation mainly refers to the GPT-like models. For the discriminative LLMs, U-BERT~\cite{qiu2021u} proposes to utilize the BERT model as embedding backbones and align the representations from the BERT model with the domain-specific data through fine-tuning. For the generative LLMs, since these models have strong zero/few-shot learning abilities, some studies propose to employ these models via prompting methods without fine-tuning. For example,~\cite{liu2023chatgpt} uses ChatGPT as a versatile recommendation model, assessing its performance across five distinct recommendation contexts. Furthermore, several studies propose to further refine the LLMs, aiming to optimize their efficacy. As an illustration, TALLRec\cite{bao2023tallrec} suggests enhancing the LLMs via recommendation-focused tuning. In this method, the input derives from user historical patterns, while the output focuses on binary feedback ("yes" or "no"). In this paper, our method mainly utilizes two different generative LLMs for recommendation, combining both of the LLM Tokens + RS paradigm and LLM as RS paradigm. Recently, several studies have also focused on employing LLMs for sequential recommendations~\cite{peng2024towards}. However, they have not considered the difficulties and challenges posed by text-rich user historical behaviors to LLMs.

\section{PROBLEM FORMULATION}
\label{sec:formulation}
Here we introduce the problem formulation of the text-rich sequential recommendation problem. Given a user $u$, we can first form the historical user behavior sequence of $u$ as $\mathcal{S} = [I_1,\dots,I_n]$, where $I_i$ is the $i$-th item the user interacted with, e.g., click, buy, read, etc., and $n$ is the length of the user behavior sequence. Each item $I$ has several types of attributes, and can be formulated as $I=[A_1,\dots,A_m]$, where $A_i$ is the $i$-th type of attribute and $m$ is the total number of attribute types. Furthermore, each attribute $A$ can be formulated in textual form as $A=[w_1,\dots,w_s]$, where $w_i$ is the $i$-th word and $s$ is the text length. Based on the above, the problem of text-rich sequential recommendation can be formulated as:

\begin{myDef}[Text-Rich Sequential Recommendation]
Given a user $u$ with the corresponding historical user behavior sequence $\mathcal{S}$, and a candidate item $I^c$, the goal of text-rich sequential recommendation is to estimate the click probability of the candidate item for user $u$, i.e., $g_u:I^c \to \mathbb{R}$.
\end{myDef}
\section{TECHNICAL DETAILS}
In this section, we will introduce our framework in detail. Specifically, we will first introduce how to get the user preference summary by the LLM-based summarizer, including the hierarchical summarization paradigm and the recurrent summarization paradigm. Then, we will elucidate the training process of the LLM-based recommender using the LoRA-based SFT method, and further demonstrate the application of the trained models for recommendation tasks.
\subsection{Hierarchical LLM-based User Preference Summarization}
\begin{figure}[tb]
    \centering
    \includegraphics[width=0.9\linewidth]{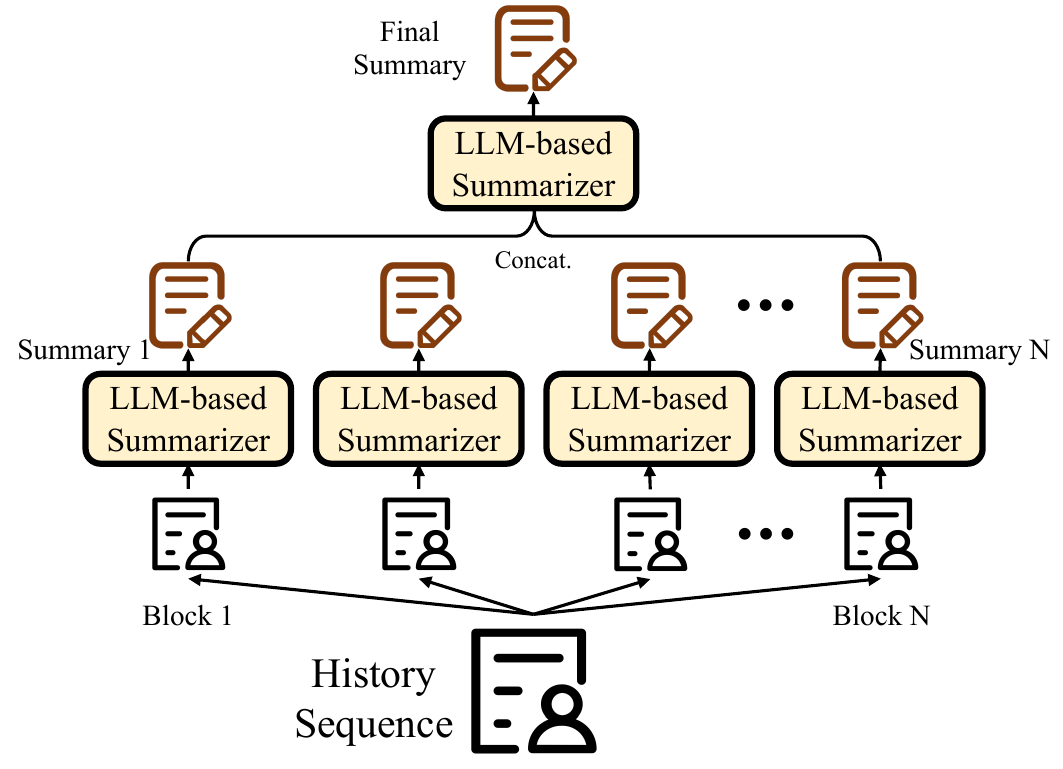}
    \vspace{-2mm}
    \caption{A schematic diagram of the hierarchical summarization paradigm.}
    \label{fig:hierarchical}
    \vspace{-2mm}
  \end{figure}
In this section, we will introduce the technical detail of the hierarchical summarization paradigm, which can be shown in Figure~\ref{fig:hierarchical}. 

\subsubsection{History Sequence Construction and Segmentation}

To harness the powerful text processing capabilities of LLMs in addressing text-rich sequential recommendation issues, given a user behavior sequence $\mathcal{S} = [I_1,\dots,I_n]$, we propose to convert $\mathcal{S}$ into a passage of text, which contains information about each item the user has interacted with. 
However, as we discussed in Section~\ref{sec:intro}, in the context of text-rich sequential recommendation, users may interact with multiple items, each containing extensive textual information. Therefore, directly processing this extensive text with LLMs can present several challenges, such as exceeding the length limitations and excessive computational resource overheads. Therefore, we propose to segment the text, ensuring that each block only contains information related to a few items, making it more manageable for further processing by the LLMs. Specific examples will be presented in the subsequent sections.

\subsubsection{Block Summarization}
\begin{figure}
\centering
\includegraphics[width=0.95\linewidth]{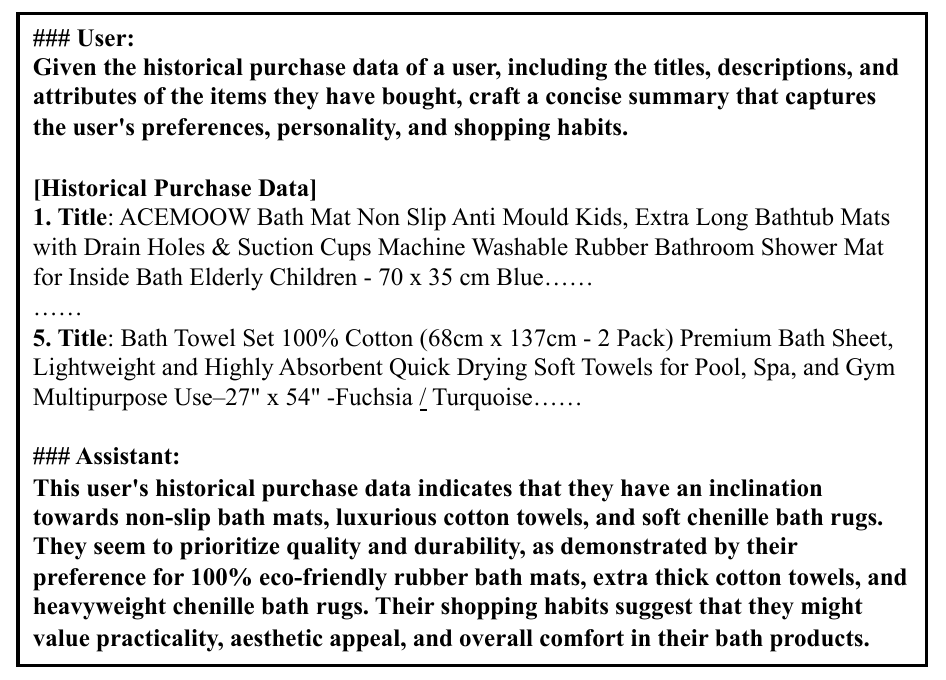}
\vspace{-2mm}
\caption{An example of block summarization on the Amazon-M2 dataset.}
\label{fig:block_summarization}
\vspace{-2mm}
\end{figure}

In the hierarchical summarization paradigm, after segmenting the text, we will subsequently summarize each text block individually, which allows us to discern the user preference within each specific time frame. For the block summarization, as mentioned in~\cite{pu2023summarization}, the zero-shot summarization capabilities of LLMs have significantly surpassed the traditional fine-tuned models, and even exceeding human performance. Therefore, in this paper, we employed the Llama-30b-instruct\footnote{https://huggingface.co/upstage/llama-30b-instruct-2048} as the summarizer, which is an LLM fine-tuned by instruction-tuning and has achieved commendable results on the Huggingface Open LLM Leaderboard\footnote{\url{https://huggingface.co/spaces/HuggingFaceH4/open_llm_leaderboard}}. Moreover, this model offers an appropriate balance between model size and performance. Figure~\ref{fig:block_summarization} illustrates how we harness the zero-shot summarization capabilities of the LLM by providing a prompt text. In this case, we input an appropriate prompt text according to the required prompt template, asking the model to summarize the user shopping preferences. The model then generates a suitable summary. We can find that, due to the inclusion of fewer items, the summaries obtained in this manner can focus on more intricate details, such as the specific materials of the products. Notably, we can achieve good versatility by modifying the prompt text. For instance, by replacing the shopping-related descriptions in the prompt with news reading-related descriptions, we can utilize the LLM to summarize the news reading preferences.

\subsubsection{Hierarchical Summarization}
\begin{figure*}
\centering
\includegraphics[width=0.95\textwidth]{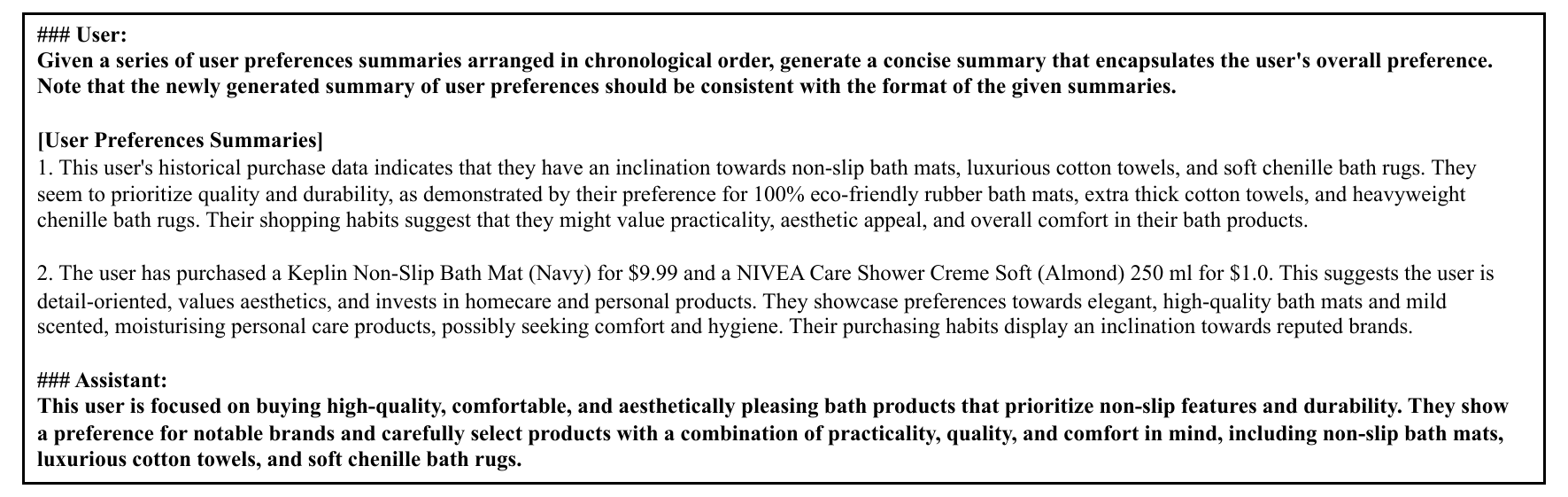}
 \vspace{-2mm}
\caption{An example of hierarchical summarization.}
\label{fig:hierarchical_summary}
\vspace{-2mm}
\end{figure*}
Under the hierarchical summarization paradigm, after obtaining the summaries for each block, we propose a hierarchical approach, which involves feeding multiple summaries into the LLM and instructing it to further summarize these summaries, ultimately yielding a comprehensive summary of the user preferences. This process bears a strong resemblance to how Convolutional Neural Network (CNN) extracts higher-level features in a layered manner. Figure~\ref{fig:hierarchical_summary} illustrates how we design an appropriate prompt to leverage the LLM for this task. 
We can find that, in contrast to the detailed focus of individual block summaries, the results derived from further summarizing multiple summaries are more abstract and general. They no longer dwell on minutiae but instead capture the overall shopping habits more effectively. This underscores the high level of abstraction and generalization capability that the hierarchical summarization paradigm can offer. It is worth noting that, although in our examples we obtained a final summary of a behavior sequence containing ten items using only two layers of summarization, we can in practice further extend the number of summarization layers, much like adding layers in a convolutional neural network. This theoretically allows us to handle behavior sequences containing information on an infinite number of items.

\subsection{Recurrent LLM-based User Preference Summarization}
\begin{figure}[tb]
    \centering
    \includegraphics[width=0.9\linewidth]{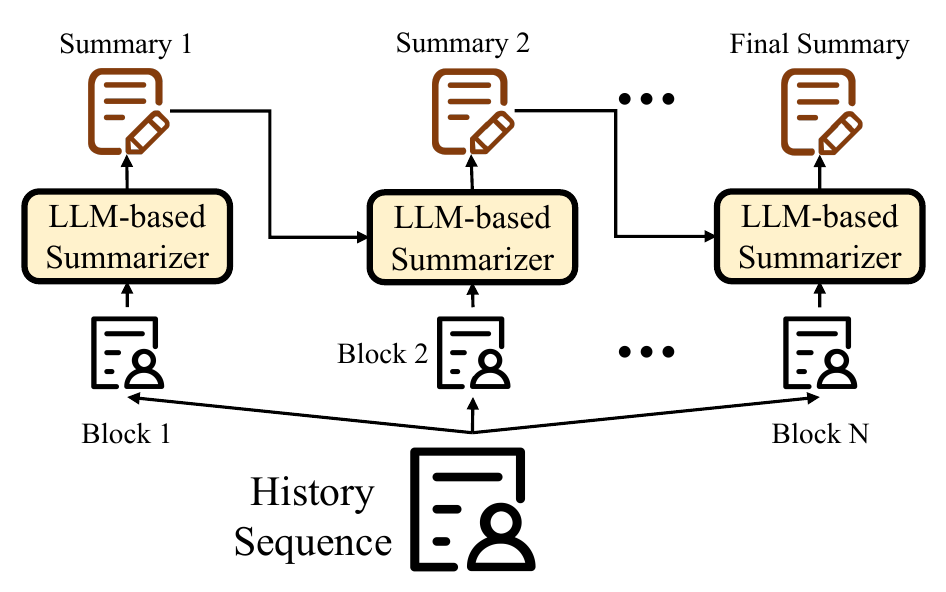}
     \vspace{-2mm}
    \caption{A schematic diagram of the recurrent summarization paradigm.}
    \label{fig:recurrent}
    \vspace{-2mm}
  \end{figure}
In this section, we will introduce the technical detail of the recurrent summarization paradigm, which can be shown in Figure~\ref{fig:recurrent}. Inspired by Recurrent Neural Networks (RNNs), the recurrent summarization paradigm operates as follows. After segmenting the user behavior sequence text into blocks, the summary of the first block is extracted. Subsequently, the summary of the preceding block and the user behavior from the next block are input into the LLM-based summarizer to produce an updated summary. This process is iteratively executed until the end of all blocks, resulting in the final user preference summary. 

\subsubsection{First Block Summarization}
\begin{figure}
\centering
\includegraphics[width=0.95\linewidth]{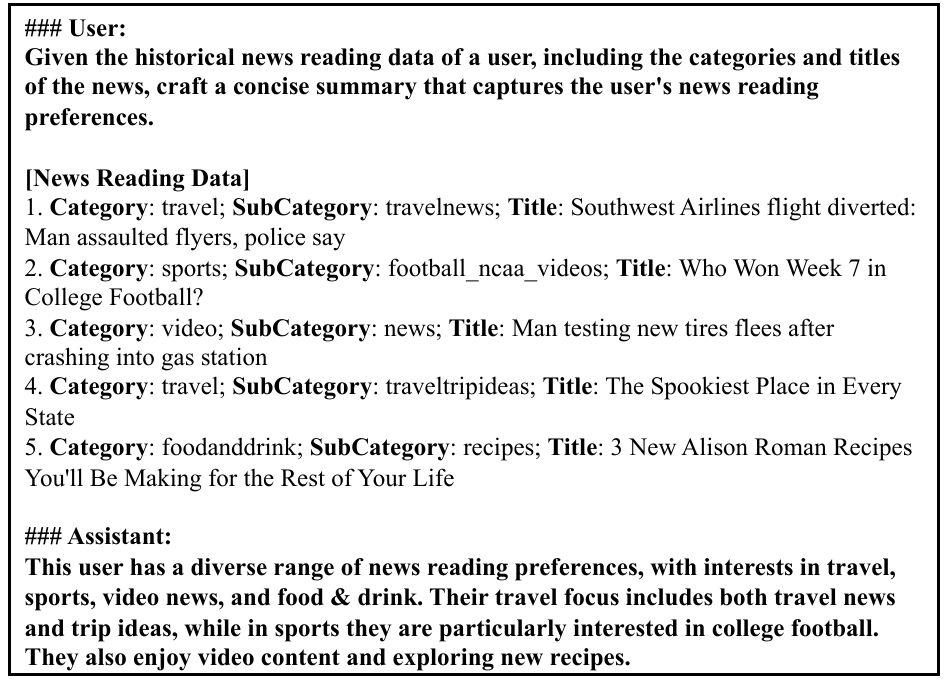}
 \vspace{-2mm}
\caption{An example of the first block summarization on the MIND dataset.}
\label{fig:first_block_MIND}
\vspace{-2mm}
\end{figure}

In the recurrent summarization paradigm, the method for summarizing the first block is essentially consistent with the approach used in the hierarchical summarization paradigm. Figure~\ref{fig:first_block_MIND} provides an example of summarizing the first block of a specific user in the MIND dataset.

\subsubsection{Recurrent Summarization}
\begin{figure*}
\centering
\includegraphics[width=0.95\textwidth]{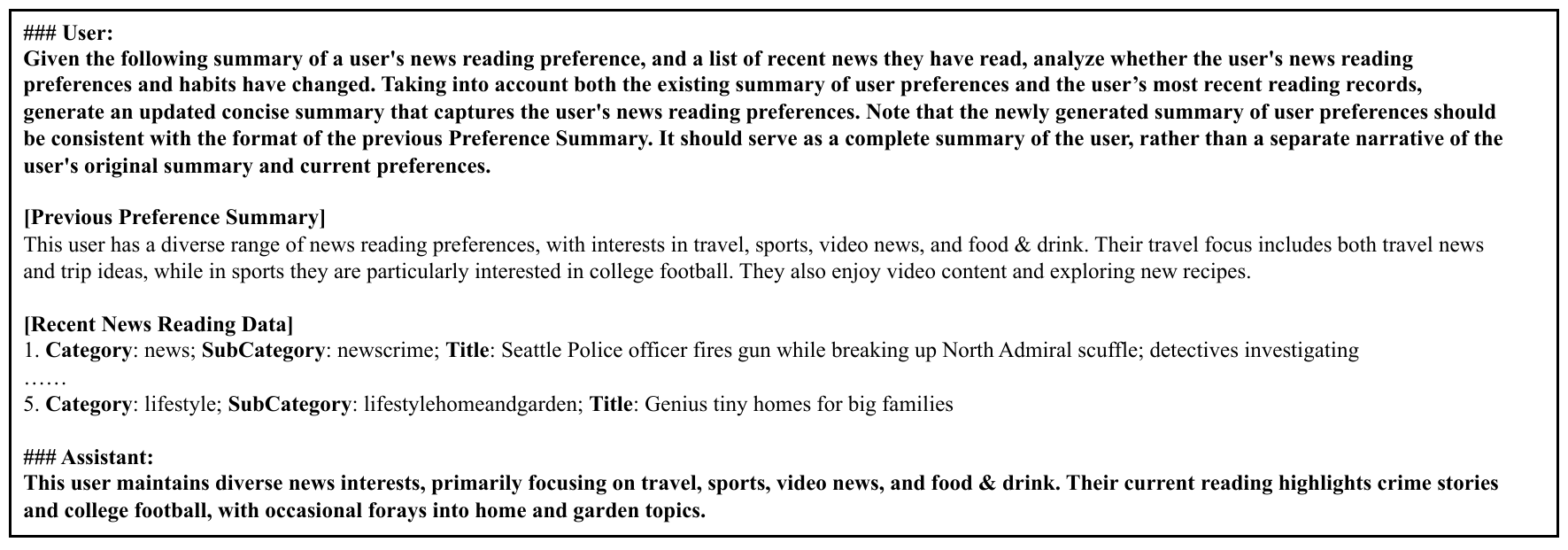}
 \vspace{-2mm}
\caption{An example of recurrent summarization.}
\label{fig:recurrent_summary}
\vspace{-2mm}
\end{figure*}
Having obtained the summary for the first block, we can proceed with a recurrent summarization to derive the final summary of user preferences. Figure~\ref{fig:recurrent_summary} demonstrates how we design an appropriate prompt text to harness the LLM for this task. It is evident that we have incorporated more detailed descriptions within the prompt text to ensure the LLM can accurately comprehend the task at hand. The output from the LLM aligns with our expectations, effectively capturing the long-term user interests while updating the summary of their short-term inclinations.

\subsection{LLM-based Recommendation}
\label{sec:LLM-based Recommendation}
\begin{figure*}
\centering
\includegraphics[width=0.95\textwidth]{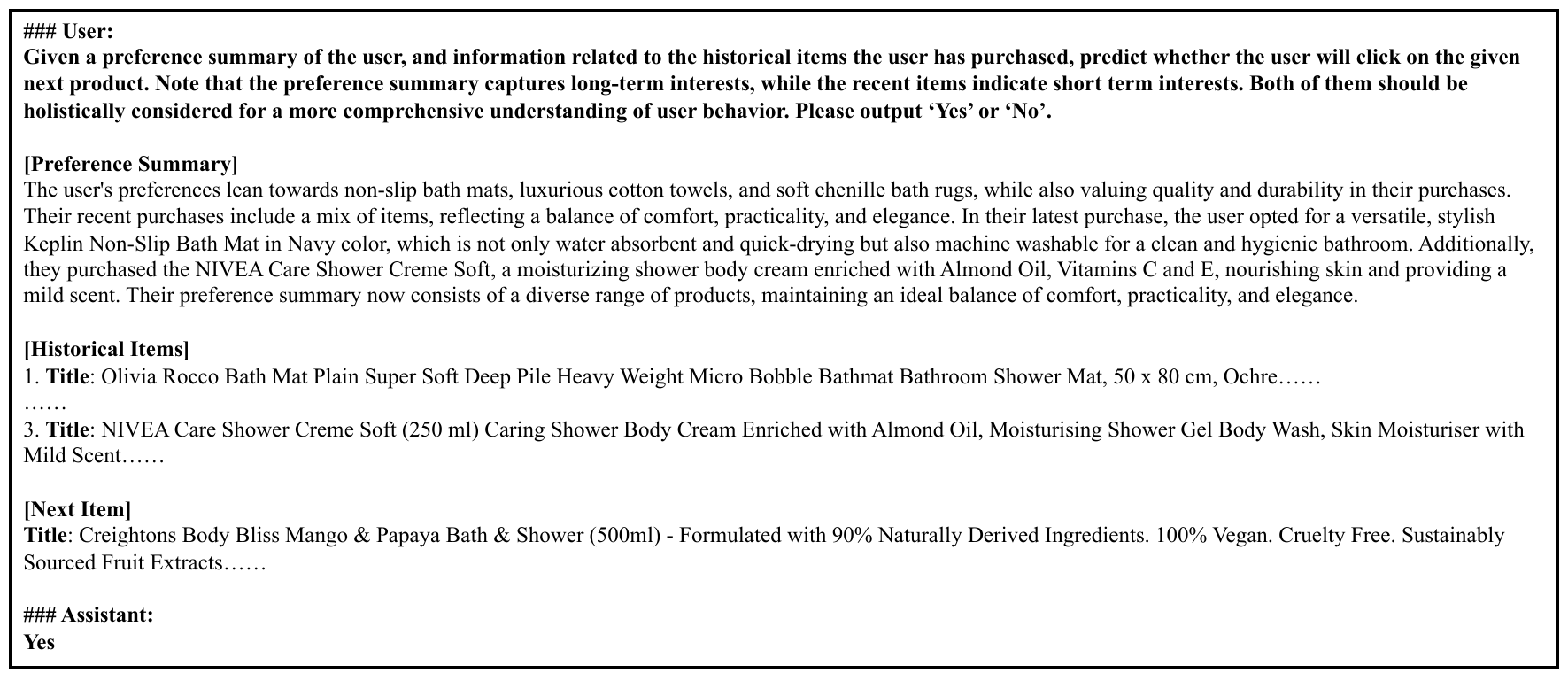}
 \vspace{-2mm}
\caption{An example of LLM-based recommendation.}
\label{fig:recommend}
\vspace{-2mm}
\end{figure*}

After getting the summary of user preferences, we can now employ an LLM-based recommender without concerns about length limitations or excessive computational overhead. We propose to train the LLM-based recommendation model using a Supervised Fine-Tuning (SFT) approach. To be specific, as illustrated in Figure~\ref{fig:recommend}, in this paper, we propose constructing a prompt text for the LLM-based recommender system composed of the following five parts:
\begin{itemize}
    \item \textbf{Recommendation Instruction}: Its role is to instruct the LLM to consider both the preference summary and the user recent behaviors to complete the recommendation task. The recommendation task is structured as an output of either "yes" or "no".
    \item \textbf{Preference Summary} This derives from the hierarchical summarization paradigm or recurrent summarization paradigm mentioned earlier, serving to represent the user long-term interests.
    \item \textbf{Recent User Behavior}: This encompasses the items the user has recently interacted with, indicating the user short-term interests.
    \item \textbf{Candidate Item Description}: This offers all textual attributes of the candidate item.
    \item \textbf{Final Answer}: This clarifies whether the user has interacted with the item or not.
\end{itemize}
Then, we use the following SFT training loss to train the LLM-based recommender as:
\begin{equation}
    \mathcal{L}_{sft} = -\sum_{i=1}^{L} \log \Pr(v_{i}|v_{<i}),
\end{equation}
where $v_i$ is the $i$-th word in the prompt text and $L$ is the length of the prompt text. The probability $\Pr(v_{i}|v_{<i})$ is calculated by the LLM model following the next-token prediction paradigm. During the training process, we utilize Low-Rank Adaptation (LoRA)~\cite{hu2021lora} for Parameter-Efficient Fine-Tuning (PEFT), which can greatly reduce the number of trainable parameters.

After the training phase is completed, during the testing phase, we remove the "yes" or "no" at the end of the prompt text. We then input this modified prompt $P$ into the large language model and obtain the probabilities predicted by the model for the next word being either "yes" or "no" as:
\begin{equation}
    p_{yes} = \Pr('yes'|P),\quad p_{no} = \Pr('no'|P).
\end{equation}
Finally, we calculate the interaction probability by using the softmax function as:
\begin{equation}
    p = \frac{\exp(p_{yes})}{\exp(p_{yes})+\exp(p_{no})}.
\end{equation}
\section{Experiments}
\label{sec:exp}
In this section, we will introduce the datasets used in this paper, the baseline methods, evaluation metrics, and experimental results.
\subsection{Dataset Description}
In this paper, we selected two open-source text-rich sequential recommendation datasets, respectively Amazon-M2 dataset~\cite{jin2023amazon} for product recommendation and MIND dataset~\cite{wu2020mind} for news recommendation. We selected records that contain a minimum of 10 and a maximum of 25 interactions. For the training set, we employed a 1:1 ratio for negative sampling, while in the validation set and test set, we utilized a 1:20 ratio for negative sampling. Detailed statistics of these two datasets are shown in Table~\ref{tab:statistics}.

\begin{table}[tb]
  \small
  \caption{Statistics of the datasets.}
  \label{tab:statistics}   
  \centering
  \begin{tabular}{l|r|r}
  \hline
  description & Amazon-M2 & MIND \\\hline
  \# of different attributes&10&4\\
  \# of positive samples in the training set&10,000&10,000\\
  \# of positive samples in the validation set&1,000&1,000\\
  \# of positive samples in the test set&1,000&1,000\\
  Avg. \# of historical user behavior sequence&13.16&16.23\\
  Avg. \# of tokens corresponding to an item&141.45&40.83\\
  \hline
  \end{tabular}
  \vspace{-4mm}
\end{table}

\begin{table*}[tb]  
    \centering  
    \caption{The performance of different models.}  
     \vspace{-2mm}
    \label{tab:main_result}  
      \begin{center}  
          \begin{tabular}{l|ccc|ccc|ccc|ccc}  
              \hline  
            ~&\multicolumn{6}{c|}{Amazon-M2}&\multicolumn{6}{c}{MIND}\\\hline
            ~&\multicolumn{3}{c|}{Recall}&\multicolumn{3}{c|}{MRR}&\multicolumn{3}{c|}{Recall}&\multicolumn{3}{c}{MRR}\\\hline
            
            ~&@3&@5&@10&@3&@5&@10&@3&@5&@10&@3&@5&@10  \\
            \midrule
            NCF&  0.8300&	0.8830&	 0.9440&  0.7328&	0.7448&	0.7529&	0.7010&	0.8030&	0.9240&	0.5523&	0.5759&	0.5926\\
            DIN& 0.7380&	0.8330&	0.9240&	0.5838&	0.6053&	0.6174&	0.7900&	0.8620&	0.9330&	0.6352&	0.6519&	0.6616 \\
            DIEN& 	0.7330&	0.8170&	0.9070&	0.5922&	0.6114&	0.6229&	0.7300&	0.8200&	0.9140&	0.6045&	0.6251&	0.6379 \\ 
            GRU4Rec& 	0.4420&	0.5590&	0.7350&	0.3355&	0.3621&	0.3855&	0.6650&	0.7970&	0.9260&	0.5305&	0.5610&	0.5787 \\
            NARM& 	0.8410&	0.8860&	0.9330&	0.7475&	0.7577&	0.7638&	0.5820&	0.7330&	0.8930&	0.4142&	0.4489&	0.4703\\
            SASRec& 0.6550&	0.7570&	0.9040&	0.4938&	0.5173&	0.5374&	0.8420&	0.8960&	0.9410&	0.7447&	0.7574&	0.7636 \\
            CORE& 	0.5230&	0.4632&	0.6450&	0.4527&	0.4632&	0.4728&	0.5170&	0.5580&	0.6370&	0.4392&	0.4488&	0.4586 \\
            TALLRec&0.8790&0.9050&0.9460&0.8585&0.8644&0.8697&0.8580&0.9020&0.9590&0.7708&0.7807&0.7885\\
            \midrule
            LLM-TRSR-Hierarchical&\textbf{0.8910}&0.9120&0.9490&0.8597&0.8643&0.8693&\textbf{0.9160}&\textbf{0.9430}&0.9750&\textbf{0.8505}&\textbf{0.8568}&\textbf{0.8611}\\
            LLM-TRSR-Recurrent&\textbf{0.8910}&\textbf{0.9130}&\textbf{0.9570}&\textbf{0.8632}&\textbf{0.8681}&\textbf{0.8737}&0.9060&0.9390&\textbf{0.9840}&0.8400&0.8475&0.8534\\
            \hline
          \end{tabular}  
      \end{center}  
    \vspace{-3mm}
    \end{table*}

\subsection{Experimental Settings}
\subsubsection{Baseline Methods and Evaluation.} 
To evaluate the performance of our model for text-rich sequential recommendation, we selected a number of state-of-art methods as baselines. 
Specifically, we first chose two traditional non-sequential recommendation methods as:
\begin{itemize}
    \item NCF~\cite{he2017neural}: NCF is a deep learning-based model for collaborative filtering. Max-pooling is used for user representation.
    \item DIN~\cite{zhou2018deep}: DIN utilizes attention mechanisms to capture the user interest from the clicked items.
\end{itemize}
Then, we chose several state-of-the-art sequential recommendation methods as:
\begin{itemize}
    \item DIEN~\cite{zhou2019deep}: DIEN adds a sequential modeling part to capture the evolution of user interest compared with the DIN model. 
    \item GRU4Rec~\cite{hidasi2015session}: GRU4Rec utilizes the GRU model for user behavior sequence modeling.
    \item CORE~\cite{hou2022core}: CORE uses a linear combination for behavior sequence modeling.
    \item NARM~\cite{li2017neural}: NARM utilizes RNNs with attention mechanisms for user behavior sequence modeling.
    \item SASRec~\cite{kang2018self}: SASRec uses self-attention combined with position embeddings for sequence modeling.
\end{itemize}
Note that for all the above baseline methods, instead of the pure ID-based paradigm, we used pre-trained BERT~\cite{devlin2018bert} model for text embedding. Finally, we chose an LLM-based sequential recommendation method as:
\begin{itemize}
    \item TALLRec~\cite{bao2023tallrec}: TALLRec proposes to leverage LLMs for recommendation by instruction tuning.
\end{itemize}
To evaluate the performance of different models, we selected Recall@K and Mean Reciprocal Rank (MRR)@K as evaluation metrics, where the value of K can be 3, 5, and 10. 

\subsubsection{Implementation Details.} 
We conducted experiments using a cluster composed of 12 Linux servers, each equipped with 8*A800 80GB GPUs. We selected Llama-30b-instruct\footnote{https://huggingface.co/upstage/llama-30b-instruct-2048} with 8-bit quantization as the summarizer and Llama-2-7b\footnote{https://huggingface.co/meta-llama/Llama-2-7b-hf} with BF16 as the recommender. We used PyTorch and TRL\footnote{https://huggingface.co/docs/trl/index} library for the SFT step and we used LoRA with the rank equal to 8. We used the AdamW~\cite{loshchilov2017decoupled} optimizer with learning rate as 1e-4 and batch size as 1 for SFT, and we set gradient accumulation steps as 64 and epoch number as 8. We also used Deepspeed~\cite{rasley2020deepspeed} with ZeRO stage as 2 for distributed training. Furthermore, we set the max length of tokens of LLMs as 2048 and the item number in a block as 5. 

\subsection{Overall Performance}
\label{sec:Overall Performance}
To demonstrate the effectiveness of our model on reciprocal recommendation, we compare LLM-TRSR with all the baseline methods, and the results are shown in Table~\ref{tab:main_result}. Note that we set the number of historical items in the prompt text for recommendation as 3, and the suffix `-Hierarchical' or `-Recurrent' indicate the paradigm through which user preference summaries are obtained. From the results, we can get the following observations:
\begin{enumerate}
    \item The performance of our model surpasses all of the baseline methods on different evaluation metrics and different datasets. This clearly proves the effectiveness of our LLM-TRSR model for text-rich sequential recommendation.
    \item Recommendation approaches based on LLMs consistently outperform traditional methods, underscoring the substantial potential of LLMs in the realm of sequential recommender systems.
    \item On the Amazon-M2 dataset, LLM-TRSR-Recurrent outperforms LLM-TRSR-Hierarchical, while the opposite holds true for the MIND dataset. This suggests that different paradigms for summarizing user preferences might be suitable for varying scenarios. For instance, the recurrent paradigm may capture the user preference transitions more effectively, whereas the hierarchical paradigm might better capture the user overarching interests.
\end{enumerate}

\subsection{Discussion on Historical Item Number}
\begin{figure}
\centering
\subfigure[The performance of LLM-TRSR-Hierarchical with different historical item number on the MIND dataset.]{
    \includegraphics[width=0.45\linewidth]{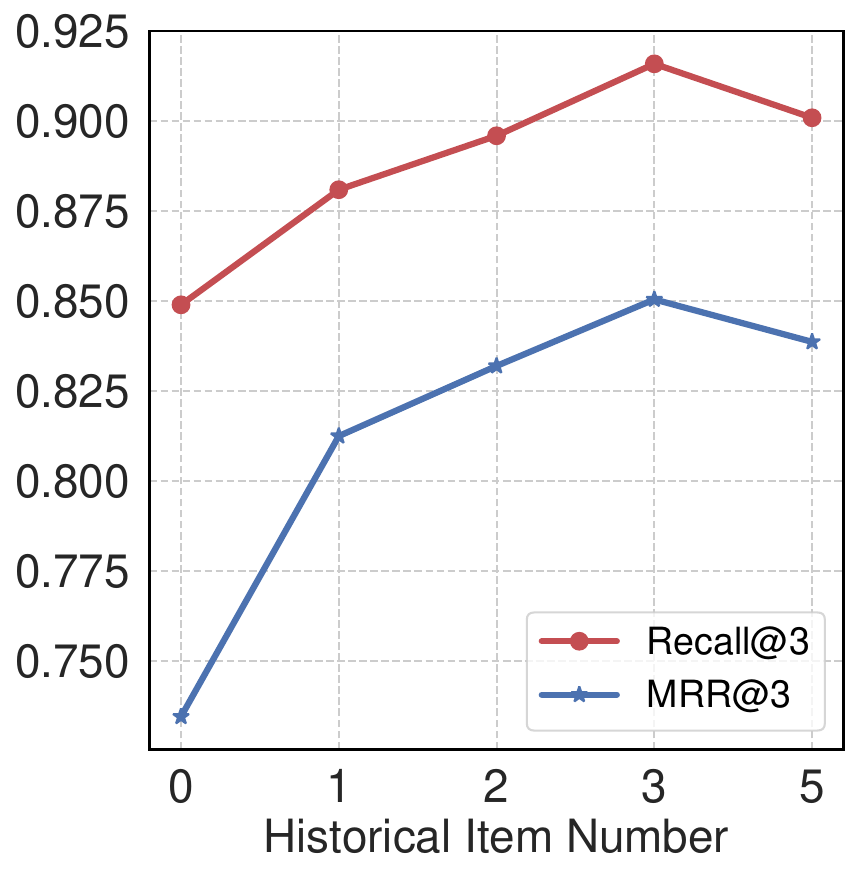}
    \label{fig:MIND_hierarchical_item_num}
     \vspace{-2mm}
}
\subfigure[The performance of LLM-TRSR-Recurrent with different historical item number on the MIND dataset.]{
    \includegraphics[width=0.45\linewidth]{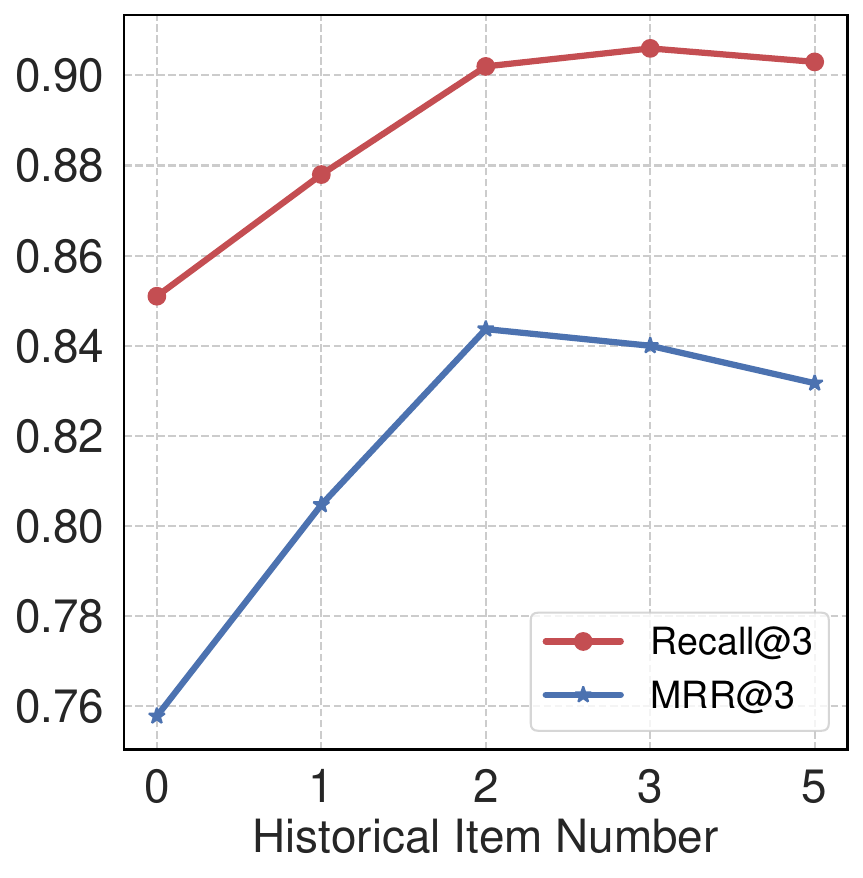}
    \label{fig:MIND_recurrent_item_num}
     \vspace{-2mm}
}
 \vspace{-4mm}
\caption{The performance of different models with different historical item number on MIND dataset.}
 \label{fig:item_num}
 \vspace{-2mm}
\end{figure}
In Section~\ref{sec:LLM-based Recommendation}, we mentioned that the prompt text fed into the LLM-based recommender includes information about items the user has historically interacted with, and in Section~\ref{sec:Overall Performance} we set this number as 3. In this section, we will explore the impact of varying numbers of historical items on the results, and the results are shown in Figure~\ref{fig:item_num}. We can find that as the number increases, the model performance initially improves and then declines, with the optimal performance occurring when the number is set to 3. This suggests that either too few or too many historical items are not conducive to enhancing the model performance. Additionally, we note that even when the number of historical items is set to 0, meaning the model recommends solely based on user preference summaries, it still achieves reasonably good performance. This underscores the effectiveness of our proposed summarization methods.

\subsection{Discussion on Parameter Size}
It is well-known that the parameter size of LLMs can significantly impact their performance. In this section, we will discuss the influence of parameter size on the framework proposed in this paper.
\subsubsection{Discussion on Parameter Size of Recommender}
\begin{figure}
\centering
\subfigure[The performance of recommenders with hierarchical summarization paradigm on the Amazon-M2 dataset.]{
    \includegraphics[width=0.45\linewidth]{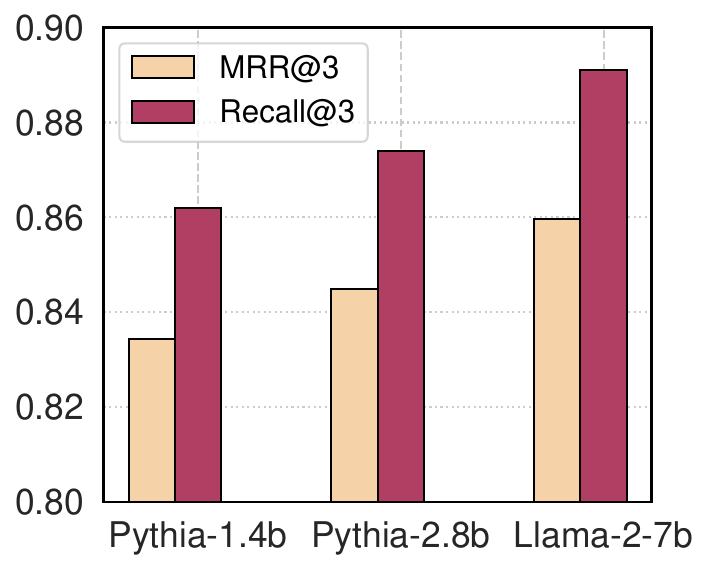}
    \label{fig:amazon_hierarchical_item_num}
     \vspace{-2mm}
}
\subfigure[The performance of recommenders with recurrent summarization paradigm on the Amazon-M2 dataset.]{
    \includegraphics[width=0.45\linewidth]{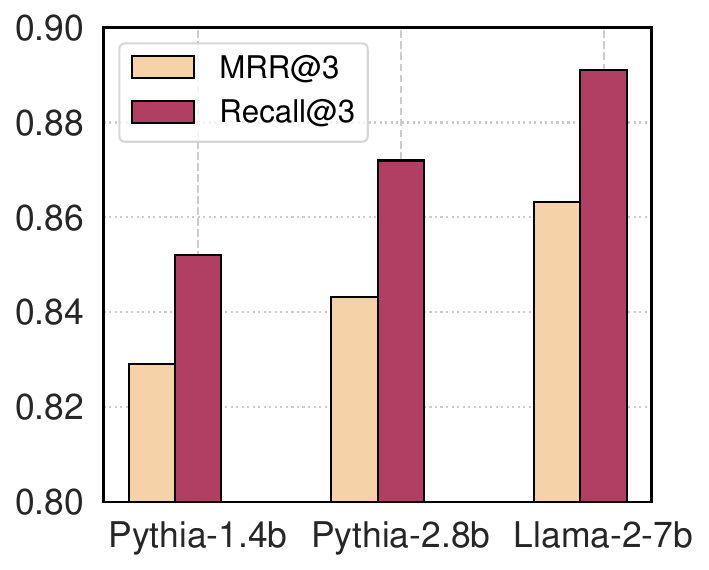}
    \label{fig:amazon_recurrent_item_num}
     \vspace{-2mm}
}
 \vspace{-2mm}
\caption{The performance of recommenders with different parameter scales on the Amazon-M2 dataset.}
 \label{fig:model_size}
  \vspace{-4mm}
\end{figure}

In this paper, we selected Llama-2-7b as the backbone model of the LLM-based recommender. To investigate the performance of recommender with varying parameter sizes, we selected models from Pythia~\cite{biderman2023pythia}, a suite
of 16 LLMs with different sizes, as comparisons. Specifically, we selected two smaller-scale models, recpectively Pythia-1.4b and Pythia-2.8b, and used them to replace the Llama-2-7b model. The results are presented in Figure~\ref{fig:model_size}. From the results, we observe that models with a larger scale generally achieve better performance. However, note that larger models also demand greater computational resources. Thus, in practical application scenarios, striking a balance between model performance and computational overhead is a matter worth considering.

\subsubsection{Discussion on Parameter Size of Summarizer}
\begin{figure}
\centering
\subfigure[The performance of summarizers with hierarchical summarization paradigm on the MIND dataset.]{
    \includegraphics[width=0.45\linewidth]{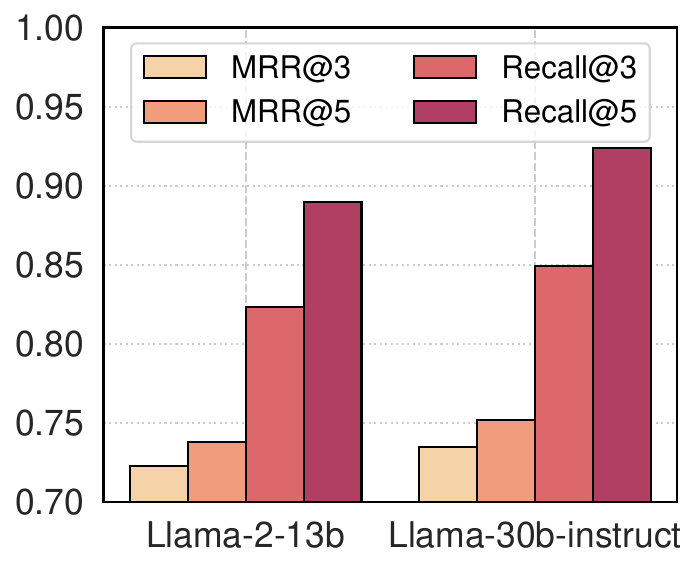}
    \label{fig:MIND_hierarchical_summary_size}
     \vspace{-2mm}
}
\subfigure[The performance of summarizers with recurrent summarization paradigm on the MIND dataset.]{
    \includegraphics[width=0.45\linewidth]{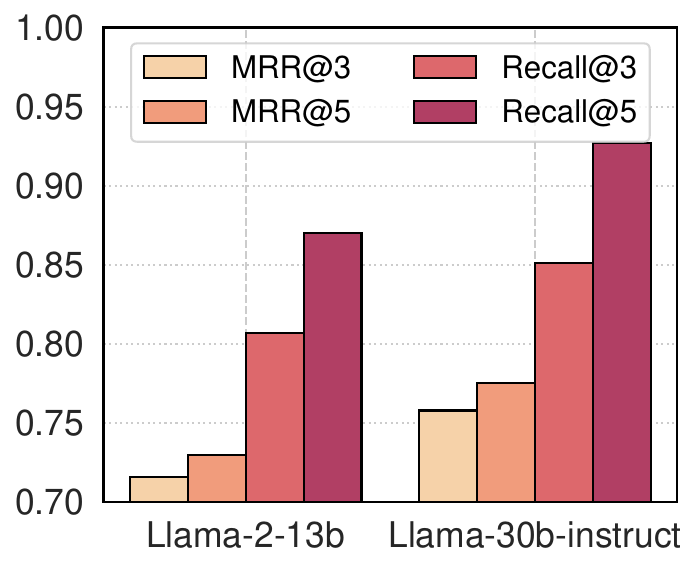}
    \label{fig:MIND_recurrent_summary_size}
     \vspace{-2mm}
}
 \vspace{-2mm}
\caption{The performance of summarizers with different parameter scales on the MIND dataset.}
 \label{fig:summarizer_size}
  \vspace{-4mm}
\end{figure}
In this paper, we employed the Llama-30b-instruct model as the summarizer, leveraging its zero-shot summarization capabilities for user preferences summarization. To investigate the differential capabilities of LLMs of varying scales in summarizing user preferences, we experimented with Llama-2-13b as the summarizer. Furthermore, to accentuate the summarization capabilities of different models, we set the historical item number in the recommendation prompt as 0. The experimental results are shown in Figure~\ref{fig:summarizer_size}. From the results, we can find that the summarization capability of Llama-30b-instruct significantly surpasses that of Llama-2-13b. Additionally, we observed that the summaries generated by Llama-2-13b were of inferior quality. Both the content and format were disorganized, making them difficult for humans to comprehend. This suggests that only LLMs with a substantial number of parameters can proficiently perform the task of zero-shot user preference summarization.

\section{CONCLUSION}
In this paper, we investigated the application of Large Language Models for Text-Rich Sequential Recommendation (LLM-TRSR). Specifically, we first proposed segmenting the user behavior sequences. Then, leveraging the zero-shot summarization capabilities of large language models, we employed an LLM-based summarizer to encapsulate user preferences. Notably, we introduced two distinct preference summarization paradigms, respectively hierarchical summarization and recurrent summarization. Subsequently, we proposed to use an LLM-based recommender for sequential recommendation tasks, with parameters being fine-tuned using Supervised Fine-Tuning (SFT). Low-Rank
Adaptation (LoRA) was also utilized for Parameter-Efficient Fine-Tuning (PEFT).  Experiments conducted on two public datasets compellingly evidenced the efficacy of our approach proposed in this paper. 

\begin{acks}
This work was partially supported by National Natural Science Foundation of China (Grant No.92370204), Guangzhou-HKUST(GZ) Joint Funding 
Program (Grant No.2023A03J0008), Education Bureau of Guangzhou Municipality, and Guangdong Science and Technology Department. This work was finished when Zhi Zheng and Wenshuo Chao worked as research intern at Career Science Lab, BOSS Zhipin.
\end{acks}

\bibliographystyle{ACM-Reference-Format}
\bibliography{main}










\end{document}